\documentclass[prd,twocolumn,showpacs,superscriptaddress,nofootinbib]{revtex4-1}
\usepackage[T1]{fontenc} 
\usepackage{amsmath}
\usepackage{amssymb}
\usepackage{amsfonts}
\usepackage{graphicx}
\usepackage{enumitem}
\usepackage{bm}
\usepackage{rotating}
\usepackage{hyperref}
\usepackage{xcolor}
\usepackage{array}
\usepackage{lmodern}
\usepackage[normalem]{ulem}
\usepackage{braket}
\usepackage{tensor}
\usepackage{mathrsfs}
\usepackage{float}

\usepackage[normalem]{ulem}

\begin{document}

\title{Post-sphaleron conversion between baryon and lepton numbers}

\author{Pei-Hong Gu}

\email{phgu@seu.edu.cn}

\affiliation{School of Physics, Jiulonghu Campus, Southeast University, Nanjing 211189, China}

\begin{abstract}

The conversion between baryon and lepton numbers has a significant impact on the mechanisms for dynamically generating the matter-antimatter asymmetry in the present universe. Under the traditional wisdom, such conversion originates from the so-called electroweak sphaleron processes. In this work we find that in association with a proper scalar and the other TeV new physics, the baryon-lepton conversion can still keep efficient after the electroweak sphalerons are no longer active. Moreover this scenario can provide a solution to the puzzle of neutron lifetime and predict some decay modes of multi-nucleons to multi-leptons.

\end{abstract}


\maketitle

\section{Introduction}


The absence of primordial antimatter in the present universe is a fundamental problem in modern physics and cosmology. This matter-antimatter asymmetry, which is equivalent to a baryon asymmetry, can be dynamically generated in an expanding universe if the particle interactions satisfy the Sakharov's three conditions \cite{sakharov1967} including the baryon number violation. In the standard model (SM), the electroweak sphaleron processes, which keep active around the temperatures $100\,\textrm{GeV} -10^{12}_{}\,\textrm{GeV}$, can violate not only the baryon number but also the lepton number by three units \cite{krs1985}. Remarkably these sphalerons conserve the baryon-minus-lepton number [$(B-L)$-number] but violate the baryon-plus-lepton number [$(B+L)$-number]. This means any successful baryogenesis mechanisms should create a net $(B-L)$-number if they work above the electroweak scale. For example, the original GUT baryogenesis \cite{yoshimura1978,ttwz1979,weinberg1979,ds1978} has been sentenced to death because it produces the same baryon and lepton numbers. Analogously, the leptogenesis \cite{fy1986} mechansim can produce a pure lepton number stored in the SM leptons before the electroweak phase transition and then the produced lepton number can be partially transferred to a baryon number stored in the SM quarks through the electroweak sphaleron processes.

In this paper, we shall show that in association with a proper scalar and the other TeV new physics, the conversion between baryon and lepton numbers can still keep efficient after the electroweak sphalerons are no longer active. Meanwhile, the same interactions can provide a solution to the puzzle of neutron lifetime and predict some decay modes of multi-nucleons to multi-leptons. The key of this scenario is that the neutron and proton decays can be kinematically suppressed even forbidden by the new scalar while the scattering processes for the baryon-lepton conversion can not be kinematically suppressed although these decays and scatterings have a common origin.

\section{Theoretical framework}


We start our demonstration from a framework of effective theory, which only extends the SM by a complex gauge-singlet scalar, i.e.
\begin{eqnarray}
\chi(1,1,0) ~~\textrm{with}~~\chi\neq \chi^\ast_{}\,.
\end{eqnarray}
Here and thereafter the brackets following the fields describe the transformations under the SM $SU(3)_c^{} \times SU(2)^{}_{L}\times U(1)_Y^{}$ gauge groups. At the renormalisable level, this gauge singlet $\chi$ in principle can be allowed to have the following terms in its potential, i.e.
\begin{eqnarray}
\label{potential}
V&\supset& \frac{1}{2}\left(\mu_2^2 +\kappa_{2}^{}\chi^\ast_{}\chi \right)\chi^2_{}+ \frac{1}{3!}\mu_3^{} \chi^3_{} + \frac{1}{4!}\kappa_{2/4}^{} \chi^4_{}+\textrm{H.c.}\,,\nonumber\\
&&
\end{eqnarray}
where the couplings $\mu_{2}^{}$ and $\mu_{3}^{}$ have a dimension of mass while the couplings $\kappa_{2}^{}$ and $\kappa_{2/4}^{}$ are dimensionless. With these quadratic, cubic and quartic terms, any global symmetries can not be conserved at all. However, some discrete symmetries are still available. Specifically, a $Z_2^{}$ discrete symmetry is allowed by the $\mu_2^{}$-term, the $\kappa_2^{}$-term and the $\kappa_{2/4}^{}$-term, a $Z_3^{}$ discrete symmetry is allowed by the $\mu_3^{}$-term, while a $Z_4^{}$ discrete symmetry is allowed by the $\kappa_{2/4}^{}$-term. Meanwhile, these discrete symmetries can be applied on the SM Higgs scalar and chiral fermions. For example, we can impose the $Z_4^{}$ discrete symmetry as below, i.e.
\begin{eqnarray}
\label{z4}
Z_4^{}:&&~~\chi(+i)\,, ~~\phi(+) \,;~~q_{Li}^{}(+i)\,,~~d_{Ri}^{}(+i)\,,~u_{Ri}^{}(+i)\,, \nonumber\\
[2mm]
&&~~l_{Li}^{}(+)\,,~~e_{Ri}^{}(+)\,,~~(i=1,2,3)\,.
\end{eqnarray}
Here $\phi$ is the SM Higgs scalar, while $q_{Li}$, $d_{Ri}^{}$, $u_{Ri}^{}$, $l_{Li}^{}$ and $e_{Ri}^{}$ are the SM three generations of chiral fermions, i.e. 
\begin{eqnarray}
\!\!&&\begin{array}{c}q_{Li}^{}(3,2, + \frac{1}{6})\end{array}\!=\!\left[\begin{array}{c}u_{Li}^{}\\
[1mm]
d_{Li}^{}\end{array}\right],\,\begin{array}{c}d_{Ri}^{}(3,1,-\frac{1}{3})\,,\end{array} \,\begin{array}{c}u_{Ri}^{}(3,1,+\frac{2}{3})\,,\end{array}\nonumber\\
[2mm]
\!\!&&\begin{array}{c}l_{Li}^{}(1,2, - \frac{1}{2})\end{array}\!=\!\left[\begin{array}{c}\nu_{Li}^{}\\
[1mm]
e_{Li}^{}\end{array}\right],\,\begin{array}{c}e_{Ri}^{}(1,1,-1)\,;\end{array}\nonumber\\
[2mm]
\!\!&&\begin{array}{c}\phi(1,2, - \frac{1}{2})\end{array}\!=\!\left[\begin{array}{c}\phi^{0}_{}\\
[1mm]
\phi_{}^{-}\end{array}\right]\,.
\end{eqnarray}
Obviously, the introduction of this $Z_4^{}$ symmetry does not affect the SM Yukawa couplings and scalar potential. It is worth mentioning that such $Z_4^{}$ symmetry can accommodate the existing schemes for generating the neutrino masses like the famous seesaw mechanism. Here the details of these schemes are not repeated for simplicity.

 \begin{figure*}
\centering
\includegraphics[scale=0.55]{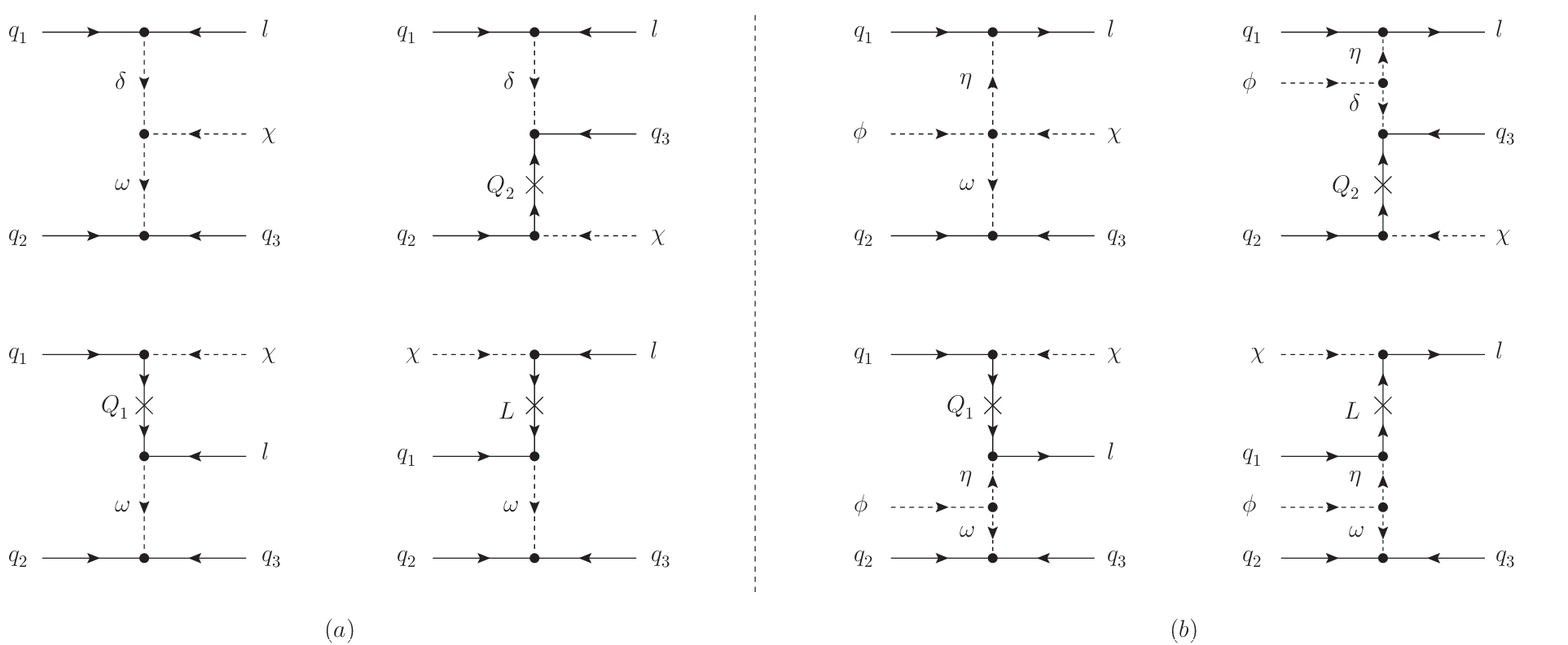} \caption{\label{eff78} The renormalizable models for the conversion between baryon and lepton numbers: (a) the $(B-L)$-conserved models, (b) the $(B+L)$-conserved models. Here the external-line fermions $(q_1^{}, q_2^{}, q_3^{}, l)$ denote the SM quarks and leptons $(q_L^{}, q_L^{}, q_L^{}, l_L^{})$, $(q_L^{}, q_L^{}, u_R^{}, e_R^{})$, $(u_R^{}, d_R^{}, q_L^{}, l_L^{})$, or $(u_R^{}, u_R^{}, d_R^{}, e_R^{})$, the internal-line scalars $\delta$, $\omega$ and $\eta$ are the corresponding color-triplets, while the internal-line fermions $Q_{1,2}^{}$ and $L$ are the corresponding vector-like quarks and leptons. It should be noted that the scalar $\eta$ is the iso-doublet, the scalars $\delta$ and $\omega$ are the iso-singlets or iso-triplets, while the fermions $Q_{1,2}^{}$ and $L$ are the iso-singlets or iso-doublets.}
\end{figure*}

In the presence of proper discrete symmetries in Eq. (\ref{potential}), the complex gauge-singlet scalar $\chi$ can be conveniently assigned not to carry any baryon or lepton numbers. Accordingly, we can construct the following dimension-7 operators to simultaneously violate the baryon and lepton numbers of one unit, i.e.
\begin{eqnarray}
\label{o1}
\mathcal{O}_{1}^{}&=& \frac{c_{ijkl}^{}}{\Lambda^3_{7}} \sum_{a=1,2,3}^{}\bar{q}_{Li}^{c}  \tau_a^{} q_{Lj}^{} \bar{q}_{Lk}^{c}  \tau_a^{} l_{Ll}^{} \chi  +\textrm{H.c.}\,,\\
[2mm]
\label{o2}
\mathcal{O}_{2}^{}&=& \frac{c_{ijkl}^{}}{\Lambda^3_{7}} \bar{q}_{Li}^{c} i\tau_2^{} q_{Lj}^{} \bar{q}_{Lk}^{c} i\tau_2^{}  l_{Ll}^{} \chi  +\textrm{H.c.}\,,\\
[2mm]
\label{o3}
\mathcal{O}_{3}^{}&=& \frac{c_{ijkl}^{}}{\Lambda^3_{7}} \bar{q}_{Li}^{c}i\tau_2^{}  q_{Lj}^{} \bar{u}_{Rk}^{c} e_{Rl}^{} \chi  +\textrm{H.c.}\,,\\
[2mm]
\label{o4}
\mathcal{O}_{4}^{}&=& \frac{c_{ijkl}^{}}{\Lambda^3_{7}} \bar{u}_{Ri}^{c} d_{Rj}^{} \bar{q}_{Lk}^{c} i\tau_2^{} l_{Ll}^{} \chi  +\textrm{H.c.}\,,\\
[2mm]
\label{o5}
\mathcal{O}_{5}^{}&=&  \frac{c_{ijkl}^{}}{\Lambda^3_{7}} \bar{u}_{Ri}^{c} d_{Rj}^{} \bar{u}_{Rk}^{c} e_{Rl}^{} \chi +\textrm{H.c.}\,.
\end{eqnarray}
Remarkably, these operators conserve the $(B-L)$-number but violate the $(B+L)$-number. Thanks to the SM Higgs scalar, we can also construct the following dimension-8 operators to violate the $(B-L)$-number but conserve the $(B+L)$-number, i.e.
\begin{eqnarray}
\label{o6}
\mathcal{O}_{6}^{}&=& \frac{c_{ijkl}^{}}{\Lambda^4_{8}} \sum_{a=1,2,3}^{}\bar{q}_{Li}^{c} \tau_a^{} q_{Lj}^{} \bar{d}_{Rk}^{c} \phi^T_{}  \tau_a^{} l_{Ll}^c  \chi  +\textrm{H.c.}\,,~~~~~\\
[2mm]
\label{o7}
\mathcal{O}_{7}^{}&=& \frac{c_{ijkl}^{}}{\Lambda^4_{8}} \bar{q}_{Li}^{c} i\tau_2^{} q_{Lj}^{} \bar{d}_{Rk}^{c} \phi^T_{} l_{Ll}^c \chi  +\textrm{H.c.}\,,\\
[2mm]
\label{o8}
\mathcal{O}_{8}^{}&=&  \frac{c_{ijkl}^{}}{\Lambda^4_{8}} \bar{u}_{Ri}^{c} d_{Rj}^{} \bar{d}_{Rk}^{c} \phi^T_{} l_{Ll}^c \chi +\textrm{H.c.}\,.
\end{eqnarray}

As shown in Fig. \ref{eff78}, the effective operators (\ref{o1}-\ref{o8}) can come from some renormalizable models with color-triplet scalars and vector-like fermions. For simplicity we here do not study these specific models and the related phenomenologies in details.

\section{Baryon-lepton conversion}


During the electroweak phase transition and the QCD phase transition, the effective operators (\ref{o1}-\ref{o8}) can result in the scattering processes as follows, 
\begin{eqnarray} 
\label{scatterings}
&&f_i^{} + f_j^{}  \longleftrightarrow f_k^c + f_l^c + \chi^\ast_{}\,,~~f_i^{} + \chi \longleftrightarrow  f_j^c + f_k^c + f_l^c  \,, ... \nonumber\\
&&
\end{eqnarray}
where the four fermions $f_{i,j,k,l}^{}$ denote the first-generation SM fermions including $(u,u,d,e)$, $(u,d,d,\nu_e^{})$ or $(u,d,d,\bar{\nu}_e^{})$. We expect these scattering processes can keep in equilibrium during the electroweak and QCD scales. For this purpose, the interaction rate of the scatterings and the Hubble constant of the universe should match the condition as below, 
\begin{eqnarray} 
\label{con1}
\Gamma(T) &>& H(T)\left |_{T>T_{D}^{}}^{}\right. ~~ \textrm{for}\nonumber\\
[2mm]
&&300\,\textrm{MeV}\sim T_\textrm{QCD}^{}< T_{D}^{}< T_{\textrm{EW}}^{}\sim 100\,\textrm{GeV}\,,~~~~
\end{eqnarray}
with $\Gamma$ being the interaction rate while $H$ being the Hubble constant. Specifically, the interaction rate can be calculated by \cite{gnrrs2003}
\begin{eqnarray}
\label{inrate}
\Gamma_{f_i^{}}^{}&=&\frac{\frac{T}{32\,\pi^4_{}}\int^{\infty}_{m_\chi^2} s^{\frac{3}{2}}_{} \sigma (s) K_1^{}\left(\frac{\sqrt{s}}{T}\right)ds }{n_{f_i^{}}^{\textrm{eq}}=\frac{2}{\pi^2_{}}T^3_{}}\,,
\end{eqnarray}
where $\sigma$ is the cross section and $K_{1}^{}$ is the Bessel function with $s$ being the squared centre of mass energy. For simplicity, the fermion masses have been ignored and the Boltzmann distribution has been adopted in the above formula. As for the Hubble constant, it can be given by \cite{kt1990}
\begin{eqnarray}
 H&=&\left(\frac{8\pi^{3}_{}g_{\ast}^{}}{90}\right)^{\frac{1}{2}}_{}\frac{T^2_{}}{M_{\textrm{Pl}}^{}}\,,
 \end{eqnarray}
where $M_{\textrm{Pl}}^{}\simeq 1.22\times 10^{19}_{}\,\textrm{GeV}$ is the Planck mass and $g_{\ast}^{}\geq 61.75$ counts the relativistic degrees of freedom slightly above the QCD scale (the SM strange, up, down, muon, electron, neutrinos, photon and gluons.).

As an example, we consider the effective operator (\ref{o4}). In this case, we can easily compute the following cross sections, 
\begin{eqnarray} 
\label{ssections}
&&\sigma(u_R^{}+d_R^{} \longrightarrow u_L^{c}+e_L^{c} +\chi^\ast_{} ) 
\nonumber\\
&=& \sigma(u_R^{}+d_R^{} \longrightarrow d_L^{c}+\nu_{eL}^{c} +\chi^\ast_{} )\nonumber\\
&=& \sigma(u_L^{}+e_L^{} \longrightarrow u_R^{c}+d_R^{c} +\chi^\ast_{} )\nonumber\\
&=& \sigma(d_L^{}+\nu_{eL}^{} \longrightarrow u_R^{c}+d_R^{c} +\chi^\ast_{} )\nonumber\\
&=& \frac{1}{ 2^{11}_{} \, \pi^3_{}} \frac{s^2}{\Lambda^6_{}}\left[1+9\frac{m_\chi^2}{s}-9 \frac{m_\chi^4}{s^2_{}} -\frac{m_\chi^6}{s^3_{}}\right.\nonumber\\
&&\left.+6\frac{m_\chi^2}{s}\left(1+\frac{m_\chi^2}{s}\right) \ln\left(\frac{m_\chi^2}{s}\right)\right]\nonumber\\
&&\textrm{with}~~\frac{1}{\Lambda^3_{}}= \frac{c_{1111}^{}}{\Lambda_7^3}\,,
\end{eqnarray}
where the SM fermion masses have been ignored and the colors have been summed over. Subsequently we can determine the interaction rate defined in Eq. (\ref{inrate}) of the scattering processes mentioned in Eq. (\ref{ssections}) to be
\begin{eqnarray}
\Gamma&=&\frac{T^7_{}}{\Lambda^6_{}}\times  \left\{\begin{array}{lll} \frac{9}{32\,\pi^5_{}}& \textrm{for} & \frac{m_\chi^{}}{T}=0\,,\\
[2mm]
6.61  \times 10^{-4}_{}&\textrm{for}& \frac{m_\chi^{}}{T}=1\,, \\
[2mm]
9.37\times 10^{-7}_{} &\textrm{for}& \frac{m_\chi^{}}{T}=10\,. \end{array}\right. 
 \end{eqnarray}
Eventually the condition (\ref{con1}) can be achieved by
\begin{eqnarray}
\label{td}
T>T_{D}^{}&=&\left\{\begin{array}{lll} 4.14\,\textrm{GeV}\left(\frac{\Lambda}{\textrm{TeV}}\right)^{1.2}_{}& \textrm{for} & \frac{m_\chi^{}}{T_{D}^{}}=0\,,\\
[2mm]
4.42\,\textrm{GeV}\left(\frac{\Lambda}{\textrm{TeV}}\right)^{1.2}_{} &\textrm{for}& \frac{m_\chi^{}}{T_{D}^{}}=1\,, \\
[2mm]
16.4\,\textrm{GeV}\left(\frac{\Lambda}{\textrm{TeV}}\right)^{1.2}_{} &\textrm{for}& \frac{m_\chi^{}}{T_{D}^{}}=10\,. \end{array}\right. 
 \end{eqnarray}

The conversion between the baryon and lepton numbers also require that the quadratic, cubic and/or quartic terms in Eq. (\ref{potential}) should be strong enough to prevent the complex scalar $\chi$ from storing any baryon and/or lepton numbers. Specifically, the $\mu_2^{}$-term leads to the oscillations $\chi\longleftrightarrow\chi^\ast_{}$; the $\mu_3^{}$-term with the inevitably quartic term, i.e. $\lambda_\chi^{}(\chi^\ast_{}\chi)^2_{}$, leads to the annihilations $\chi\chi^\ast_{}\longleftrightarrow \chi\chi\chi$, $\chi\chi^\ast_{}\longleftrightarrow \chi^\ast_{}\chi^\ast_{}\chi^\ast_{}$, $\chi\chi\longleftrightarrow \chi\chi^\ast_{}\chi^\ast_{}$ and $\chi^\ast_{}\chi^\ast_{}\longleftrightarrow \chi\chi\chi^\ast_{}$; the $\kappa_2^{}$-term leads to the annihilations $\chi\chi^\ast_{}\longleftrightarrow \chi\chi$ and $\chi\chi^\ast_{}\longleftrightarrow\chi^\ast_{}\chi^\ast_{}$; while the $\kappa_{2/4}^{}$-term leads to the annihilations $\chi\chi \longleftrightarrow \chi^\ast_{}\chi^\ast_{}$. These oscillating and/or annihilating processes should keep in equilibrium before the scattering processes (\ref{scatterings}) are frozen out. For demonstration, we consider the $Z_4^{}$ discrete symmetry and then the $\kappa_{2/4}$-term. We hence compute the cross section,  
\begin{eqnarray}
\sigma(\chi\chi \longleftrightarrow \chi^\ast_{}\chi^\ast_{}) = \frac{\kappa_{2/4}^2}{32\,\pi}\frac{1}{s}\,,
\end{eqnarray}
and then the interaction rate,
\begin{eqnarray}
\Gamma_\chi^{}&=&\frac{\frac{T}{32\,\pi^4_{}}\int^{\infty}_{4m_\chi^2} \sqrt{s}(s-4m_\chi^2) \sigma(s) K_1^{}\left(\frac{\sqrt{s}}{T}\right) ds }{n^{\textrm{eq}}_{\chi}=\frac{1}{\pi^2_{}}m_\chi^2 T K_{2}^{}\left(\frac{m_\chi^{}}{T}\right)} \nonumber\\
&=&\kappa_{2/4}^2 T \times  \left\{\begin{array}{lll} \frac{1}{512\,\pi^3_{}} & \textrm{for} & \frac{m_\chi^{}}{T}=0\,,\\
[2mm]
2.17  \times 10^{-5}_{}&\textrm{for}& \frac{m_\chi^{}}{T}=1\,, \\
[2mm]
6.89\times 10^{-9}_{} &\textrm{for}& \frac{m_\chi^{}}{T}=10\,. \end{array}\right. 
\end{eqnarray}
Again, the Boltzmann distribution has been adopted for simplicity. By comparing the above interaction rate to the Hubble constant, we find the annihilating process $\chi\chi \longleftrightarrow \chi^\ast_{}\chi^\ast_{}$ can keep in equilibrium below the temperature as follows,
\begin{eqnarray}
\label{te}
T<T_{E}^{}&=&\kappa_{2/4}^2  \times  \left\{\begin{array}{lll} 5.62\times 10^{13}_{}\,\textrm{GeV}& \textrm{for} & \frac{m_\chi^{}}{T}=0\,,\\
[2mm]
1.94\times 10^{13}_{}\,\textrm{GeV}&\textrm{for}& \frac{m_\chi^{}}{T}=1\,, \\
[2mm]
6.15 \times 10^{9}_{}\,\textrm{GeV}&\textrm{for}& \frac{m_\chi^{}}{T}=10\,. \end{array}\right. 
\end{eqnarray}

Now we have confirmed that the temperature $T_D^{}$ in Eq. (\ref{td}) can be much lower than the temperature $T_E^{}$ in Eq. (\ref{te}) even if the coupling $\kappa_{2/4}^{}$ of the complex gauge-singlet scalar $\chi$ takes a tiny value. This means that the baryon and lepton numbers can be converted from each other after the electroweak sphaleron processes stop working. Similarly, the other couplings $\mu_2^{}$, $\mu_3^{}$ and $\kappa_2^{}$ can also realize the expected oscillations and/or annihilations even if their values are very small. Ones may wonder why we do not consider the simplest case with the scalar $\chi$ being real rather than complex. If the scalar $\chi$ is real, it can automatically carry no baryon or lepton numbers. However, we will show shortly that this case is additionally constrained by the di-nucleon decays.

\section{ Nucleon decays}

 \begin{figure}
\centering
\includegraphics[scale=0.35]{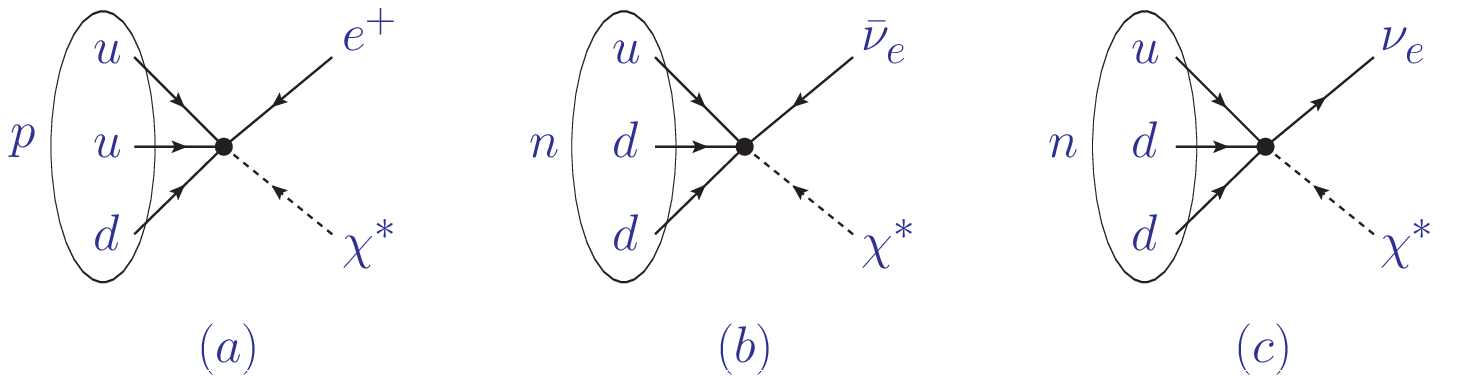} \caption{\label{npdecays} The proton and neutron decays: (a) $p\rightarrow e^{+}_{}+\chi^\ast_{}$; (b) $n\rightarrow \bar{\nu}_{e}^{} +\chi^\ast_{}$, (c) $n\rightarrow \nu_{e}^{} +\chi^\ast_{}$.}
\end{figure}


It is easy to understand that the effective operators (\ref{o1}-\ref{o8}) can result in certain proton and/or neutron decay modes as long as the kinematics is allowed. The relevant Feynman diagrams are shown in Fig. \ref{npdecays}. As an example, we still focus on the effective operator (\ref{o4}). In this case, the effective Lagrangian contributing to the exotic nucleon decays is given by,
\begin{eqnarray}
\label{effectivepndecay}
\mathcal{L}^{\textrm{eff}}_{}= \frac{\alpha}{\Lambda^3_{}} \bar{e}_{L}^c  p\chi  + \frac{\alpha}{\Lambda^3_{}} \bar{\nu}_{eL}^c n \chi  + \textrm{H.c.}\,,
\end{eqnarray}
where the parameter $\alpha = 0.0144\, \textrm{GeV}^3_{}$ is from the Lattice QCD calculations \cite{aiss2017}.

The exotic proton decay from Eq. (\ref{effectivepndecay}) should be completely forbidden or highly suppressed to fulfil the null results from experimental searches. Unless the effective cutoff $\Lambda$ is above the GUT scale, the complex gauge-singlet scalar $\chi$ should have the mass range as below, 
\begin{eqnarray}
\Gamma(p \rightarrow e_{}^{+} +\chi) \!&\equiv &\!0~~\textrm{for}~~m_\chi^{}\geq m_p^{} - m_e^{}=937.761\,\textrm{MeV}\,,\nonumber\\
[2mm]
\Gamma(p \rightarrow e_{}^{+} +\chi)\!&\gtrapprox&\! 0 ~~\textrm{for}~~m_\chi^{} \lessapprox  m_p^{} - m_e^{}=937.761\,\textrm{MeV}\,.\nonumber\\
&&\!
\end{eqnarray}

On the other hand, the precise lifetime of neutron is still an open question \cite{wg2011,gg2016}. There are two qualitatively different types of direct neutron lifetime measurements: bottle and beam experiments. In the bottle method, the ultracold neutrons are stored in a container for a time comparable to the neutron lifetime. The remaining neutrons are counted and then the neutron lifetime is extracted. In the beam method, the neutrons in a beam and the protons from these neutron decays are both counted, and then the neutron lifetime is obtained from the decay rate. Currently there is a $4\,\sigma$ discrepancy in these two neutron lifetime measurements \cite{mbmpf1993,serebrov2005,pvsg2010,spkmd2012,arzumanov2015,serebrov2017,pattie2017,bd1996,yue2013}, i.e.
\begin{eqnarray}
\label{discrepancy}
\left.\begin{array}{r} \tau_{n}^{\textrm{bottle}} = 879.6 \pm 0.6 \,\textrm{s}\\
[2mm]
\tau_{n}^{\textrm{beam}} = 888.0 \pm 2.0 \,\textrm{s}\end{array} 
\right\} \Longrightarrow  \Delta \tau_n^{} = 8.4\,\textrm{s} \,.
\end{eqnarray}
This means that either there should be an uncontrolled systematic error in one of the measurement methods, or there should be a theoretical reason why the two methods give different results. Remarkably, the exotic neutron decay from Eq. (\ref{effectivepndecay}) can provide the latter possibility \cite{fg2018,khatibi2023,llmw2023}. Actually we have 
\begin{eqnarray}
\label{endw}
\Gamma(n \rightarrow \bar{\nu}_{e}^{} +\chi) \!&=&\! \frac{1}{16\pi} \frac{\alpha^2_{}}{\Lambda^6_{}}\left[1-\frac{m_\chi^2}{m_n^2} \right]^2_{} m_n^{} ~~\textrm{for}~~m_\chi^{}< m_n^{}\,.\nonumber\\
\!&&\!
\end{eqnarray}
Moreover, this neutron decay generally could trigger the nuclear transitions from $(Z, A)$ to $(Z, A-1)$. Specifically, the requirement of $^{9}\textrm{Be}$ stability puts the most stringent constraint \cite{ahmed2003,araki2005,takhistov2015}, i.e.  
\begin{eqnarray}
\label{mchi}
937.900\,\textrm{MeV} < m_\chi^{} < m_n^{}= 939.565\,\textrm{MeV}\,.
\end{eqnarray}

The discrepancy (\ref{discrepancy}) between the values measured in bottle and beam experiments corresponds to the decay width difference as below \cite{fg2018},
 \begin{eqnarray}
 \label{discrepancy2} 
 \Delta \Gamma = \Gamma^{\textrm{bottle}}_{}- \Gamma^{\textrm{beam}}_{}=7.1\times 10^{-30}_{}\,\textrm{GeV}\,. 
\end{eqnarray}
Since the neutron decay in beam experiments is observed by detecting the produced protons, its measured lifetime now should be the same as the SM one dominated by the $\beta$ decay. To resolve the neutron decay anomaly (\ref{discrepancy}), the exotic decay width (\ref{endw}) should be equal to the measured width difference (\ref{discrepancy2}), i.e.
\begin{eqnarray}
\Gamma(n \rightarrow \bar{\nu}_{e}^{} +\chi) \!&=& \Delta \Gamma = 7.1\times 10^{-30}_{}\,\textrm{GeV} \,.
\end{eqnarray}
For this purpose, we can take a proper parameter choice such as 
\begin{eqnarray}
m_\chi^{}= 938\,\textrm{MeV}\,,~~\Lambda= 1.35\,\textrm{TeV} \,,
\end{eqnarray}
in Eq. (\ref{endw}).

 \begin{figure*}
\centering
\includegraphics[scale=0.35]{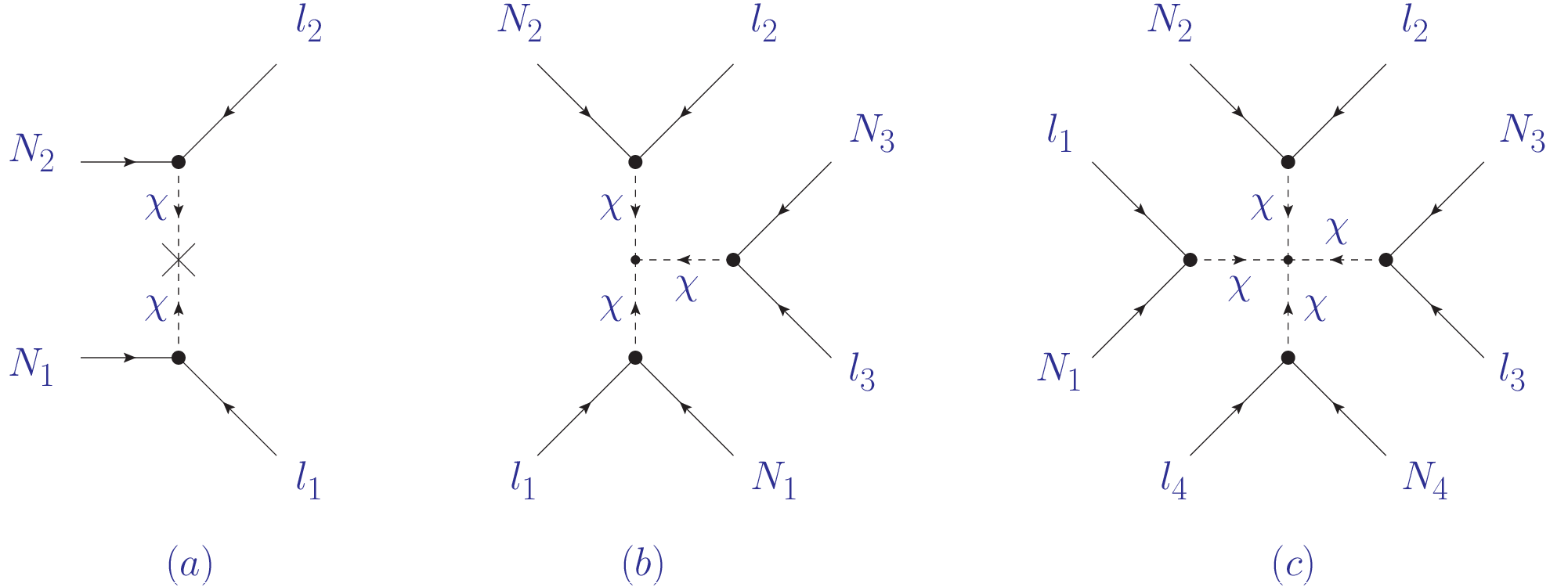} \caption{\label{4n4l} The decays of multi-nucleons to multi-leptons. Here the external-line fermions $(N_{a}^{}, l_{a}^{})$, $(a=1,2,3,4)$, denote the nucleons and leptons $(p, e^+{})$, $(n,\bar{\nu}_{e}^{})$ or $(n,\nu_{e}^{})$. Specifically, a $Z_2^{}$ discrete symmetry allows the diagrams (a) and (c), a $Z_3^{}$ discrete symmetry allows the diagram (b), while a $Z_4^{}$ discrete symmetry allows the diagram (c). The decay modes with external-line complex scalar $\chi$, i.e. $N_1^{}N_2^{}\longrightarrow l_1^{c} l_2^{c} \chi^\ast_{}\,,~l_1^{c} l_2^{c} \chi^\ast_{} \chi\,,~l_1^{c} l_2^{c} \chi^\ast_{}\chi^\ast_{}\,,~l_1^{c} l_2^{c} \chi\chi$ and $N_1^{}N_2^{}N_3^{}\longrightarrow l_1^{c} l_2^{c} l_3^{c} \chi^\ast_{}$, can take place when the kinematics is allowed. The relevant diagrams are not shown for simplicity.}
\end{figure*}

The effective operators (\ref{o1}-\ref{o8}) should be additionally constrained by the decays of multi-nucleons to multi-leptons due to the quadratic, cubic and/or quartic terms in Eq. (\ref{potential}), even if their predicted proton and neutron decays are not allowed by kinematics. The relevant Feynman diagrams are shown in Fig. \ref{4n4l}. For example, we consider the decays of di-nucleons to di-leptons \cite{afw2013,gs2020,gs2019,gs2020-2,bkl2015,girmohanta2025,hm2021,babu2020}. Without exactly calculating the phase space and the nuclear matrix elements, we can roughly estimate the decay width to be much smaller than the magnitude as follows, 
\begin{eqnarray}
\Gamma\left(2N\rightarrow 2l\right)&\ll &\left(2m_N^{}\right) \left(\frac{\alpha}{\Lambda^3}\right)^4_{} \nonumber\\
[2mm]
&&\times \left[\left(\frac{q^2_{}-m_N^{2}}{
q^2_{}-m_{\chi_R^{}}^{2}}\right)^{2}_{} - \left(\frac{q^2_{}-m_N^{2}}{
q^2_{}-m_{\chi_I^{}}^{2}}\right)^{2}_{}\right] \nonumber\\
[2mm]
&\simeq&  \frac{4\mu_2^2 m_N^5}{m_\chi^6} \left(\frac{\alpha}{\Lambda^3}\right)^4_{}\nonumber\\
[2mm]
&&\textrm{with}~~m_{\chi_R^{}}^2-m_{\chi_I^{}}^2=\mu^2_{2}\ll m_\chi^2 \,.
\end{eqnarray}
Here $\chi_{R,I}^{}$ are the real and imaginary parts of the complex scalar $\chi$, while $q \simeq 100\,\textrm{MeV}$ is the transfer moment in nuclei. Accordingly, the lifetime can be sensitive to the experimental limits \cite{berger1991,araki2005,gustafson2015,sussman2018,takhistov2015} for a proper parameter choice, i.e.
\begin{eqnarray}
\tau=\frac{1}{\Gamma}\gg 10^{29}_{}\,\textrm{yrs}~\times \left( \frac{\Lambda}{\textrm{TeV}}\right)^{12}_{}  \left(\frac{\textrm{eV}}{\mu_2^{}}\right)^4_{} \left( \frac{m_\chi^{}}{m_p^{}}\right)^{6}_{}\,.~~~~
\end{eqnarray}
Obviously, the di-nucleon decays would not be suppressed if the scalar $\chi$ was real rather than complex. We also check the decays of quad-nucleons to quad-leptons. In this case, the decay width should be much smaller than the following magnitude, i.e.
\begin{eqnarray}
\Gamma\left(4N\rightarrow 4l\right)&\ll &\kappa^2_{2,4} \left(\frac{\alpha}{\Lambda^3}\right)^8_{} \left(\frac{q^2_{}-m_N^{2}}{
q^2_{}-m_\chi^{2}}\right)^{8}_{}  \left(4m_N^{}\right)\,.~~
\end{eqnarray}
In consequence, the lifetime can be far greater than the cosmic lifespan, i.e.
\begin{eqnarray}
\tau=\frac{1}{\Gamma}\gg 3\times 10^{54}_{}\,\textrm{yrs}~\times\frac{1}{\kappa_{2,4}^2} \left( \frac{\Lambda}{\textrm{TeV}}\right)^{24}_{}  \left( \frac{m_\chi^{}}{m_p^{}}\right)^{16}_{}\,. ~~~~
\end{eqnarray}

If the complex gauge-singlet scalar $\chi$ is a dark matter particle, its mass should be in a narrow rage determined by Eq. (\ref{mchi}). The effective operators (\ref{o1}-\ref{o8}) can accommodate a freeze-in mechanism to give this dark matter scalar a right relic density \cite{khatibi2023}. We can also consider the traditional freeze-out mechanism by introducing a dark Higgs scalar to spontaneously break a dark $U(1)$ gauge symmetry during the GeV and MeV scales. Then two dark matter scalars can efficiently annihilate into two dark Higgs scalars, benefited from the Higgs portal interaction. Clearly, the realization of this scenario implies that the dark symmetry breaking should not take place before the dark matter annihilation is frozen out. Alternatively, two dark matter scalars can efficiently annihilate into two dark Higgs bosons and into two dark photons. As for the scattering of dark matter off nuclei can be mediated by the dark photon. For simplicity, the details of dark matter are not studied here.

Furthermore, we can introduce three SM-singlet right-handed neutrinos $\nu_{Ri}^{}(1,1,0)$ with zero or small Majorana masses and then construct the following effective operators,
\begin{eqnarray}
\label{o9}
\mathcal{O}_{9}^{}&=&\frac{c_{ijkl}^{}}{\Lambda^3_{7}} \bar{q}_{Li}^{c}i\tau_2^{}  q_{Lj}^{} \bar{d}_{Rk}^{c} \nu_{Rl}^{} \chi  +\textrm{H.c.}\,,\\
[2mm]
\label{o10}
\mathcal{O}_{10}^{}&=& \frac{c_{ijkl}^{}}{\Lambda^3_{7}} \bar{u}_{Ri}^{c} d_{Rj}^{} \bar{d}_{Rk}^{c} \nu_{Rl}^{} \chi +\textrm{H.c.}\,, \\
[2mm]
\label{o11}
\mathcal{O}_{11}^{}&=&\frac{c_{ijkl}^{}}{\Lambda^4_{8}} \bar{q}_{Li}^{c} i\tau_2^{} q_{Lj}^{} \bar{q}_{Lk}^{c} i\tau_2^{} \phi \nu_{Rl}^c \chi  +\textrm{H.c.}\,,\\
[2mm]
\label{o12}
\mathcal{O}_{12}^{}&=& \frac{c_{ijkl}^{}}{\Lambda^4_{8}} \bar{u}_{Ri}^{c} d_{Rj}^{} \bar{q}_{Lk}^{c} i\tau_2^{} \phi \nu_{Rl}^c \chi  +\textrm{H.c.}\,,
\end{eqnarray}
to enforce some exotic neutron decays although these operators do not achieve the post-sphaleron conversion between the SM baryon and lepton numbers. However, such right-handed neutrinos can perform some interesting phenomenologies in the framework of renormalizable models. For example, the right-handed neutrinos could appear in the final sates of the leptoquark decays and hence could modify the branching fraction of the same leptoquark decaying into a charged lepton. Consequently the low limits on the leptoquark masses can be relaxed significantly \cite{takahashi2026} if the leptoquarks are responsible for mediating the effective operators (\ref{o3}) and (\ref{o5}).

\section{Conclusion }


In the present work we have proposed a mechanism to achieve a post-sphaleron conversion between baryon and lepton numbers. Specifically, we introduce a complex gauge-singlet scalar without carrying any baryon or lepton numbers to construct possible $(B-L)$-number-conserving or $(B+L)$-number-conserving operators for simultaneously violating the baryon number and the lepton number by one unit. We also show the renormalizable models by extending proper color-triplet scalars, vector-like quarks and/or vector-like leptons. When the effective cutoff is taken at the TeV scale, the high-dimensional operators of the new scalar to three SM quarks and one SM lepton can induce some scattering processes to enforce an efficient baryon-lepton conversion after the electroweak phase transition. On the other hand, the same interactions can accommodate a kinematically forbidden proton decay, provide a solution to the puzzle of neutron lifetime, and predict some decay modes of multi-nucleons to multi-leptons.

\textbf{Acknowledgement}: This work was supported in part by the National Natural Science Foundation of China under Grant No. 12175038.

\end{document}